\documentclass[twocolumn,amsmath,pra,nobibnotes,showpacs,superscriptaddress]{revtex4}
\usepackage{graphicx}

\begin{document}
\title{Tomographic test of Bell's inequality for a time-delocalized single photon}

\author{Milena D'Angelo}
\affiliation{LENS, Via Nello Carrara 1, 50019 Sesto Fiorentino, Florence, Italy}

\author{Alessandro Zavatta}
\affiliation{Istituto Nazionale di Ottica Applicata, L.go E. Fermi, 6, I-50125, Florence, Italy}

\author{Valentina Parigi}
\affiliation{LENS, Via Nello Carrara 1, 50019 Sesto Fiorentino, Florence, Italy}\affiliation {Department
of Physics, University of Florence, I-50019 Sesto Fiorentino, Florence, Italy}

\author{Marco Bellini}
\email{bellini@inoa.it} \affiliation{LENS, Via Nello Carrara 1, 50019 Sesto Fiorentino, Florence,
Italy}\affiliation{Istituto Nazionale di Ottica Applicata, L.go E. Fermi, 6, I-50125, Florence, Italy}

\date{\today}

\begin{abstract}
Time-domain balanced homodyne detection is performed on two well-separated temporal modes sharing a
single photon. The reconstructed density matrix of the two-mode system is used to prove and quantify its
entangled nature, while the Wigner function is employed for an innovative tomographic test of Bell's
inequality based on the theoretical proposal by Banaszek and Wodkiewicz (Phys. Rev. Lett. 82, 2009 (1999)). Provided some
auxiliary assumptions are made, a clear violation of Banaszek-Bell's inequality is found.
\end{abstract}


\maketitle

\section{Introduction}\label{sec:intro}

The concept of entanglement, as first introduced by Schr{\"{o}}dinger \cite{schro} and discussed by
Einstein, Podolsky, and Rosen  \cite{epr} in 1935, has historically been associated with systems of two
or more quanta. Its counterintuitive consequences found a mathematical formulation in the work of Bell
\cite{bell} of 1965: pairs of quanta entangled in discrete variables (i.e., Bohm-like entangled systems
\cite{bohm}) may give rise to purely quantum nonlocal correlations. Unfortunately, no loophole-free test
of Bell's inequality has been realized so far \cite{note_assumpt}. Besides their relevance in
fundamental physics, these phenomena have attracted much attention due to their usefulness in quantum
information technology. Extravagant but promising protocols such as quantum teleportation, quantum
cryptography, and quantum computation have been proposed and experimentally verified (see, e.g.,
\cite{qu_infor} and references therein).

In the early nineties, increasing attention has been given to a new perspective of quantum entanglement.
The concept of entanglement has indeed been extended to any system containing a fixed number
($N=1,2,..$) of photons, provided at least one pair of spatiotemporal modes is involved. Then, even for
$N=1$, one may talk of ``single-photon n-mode entanglement", provided $n \geq 2$. The first steps in
this direction were taken by Tan, Walls, and Collet \cite{tan_prl91} (TWC): when a single photon
impinges on a beam-splitter whose outputs are described by the spatial modes $A$ and $B$, the emerging
single photon is described by the state $|\psi \rangle=\alpha |1\rangle_A |0\rangle_B + \beta
|0\rangle_A |1\rangle_B$, where $|0\rangle$ is the vacuum, $|1\rangle$ is a single-photon Fock state,
and $\alpha$ and $\beta$ are complex amplitudes such that $|\alpha|^2+|\beta|^2=1$. In this perspective,
a single photon, with its presence or absence from a given spatiotemporal mode, may represent a tool for
entangling two (or more) well-separated field modes. In this case, an experimental test of nonlocality
would consist in performing simultaneous measurements on the separate field modes ``sharing'' the single
photon and check if their mutual correlations violate a Bell's inequality.

Intensive studies and debates have been dedicated to the meaningfulness of extending both the concepts
of entanglement and nonlocality to TWC-type single-photon two-mode systems (see, e.g.,
\cite{nonloc_1phot,hardy_prl94,lvovsky_prl04_2,banaszek,vanEnk,powlowsky,noi,hessmo_prl04}). Quite
recently, Babichev, et al.\cite{lvovsky_prl04_2} and Zavatta, et al.\cite{noi} have experimentally
characterized a source of two-mode spatially and temporally delocalized single photons, respectively, by
employing time-domain homodyne tomography \cite{leonhardt} and reconstructing both density matrix and
Wigner function of the measured system. In this respect, it is worth reminding that the density matrix
or Wigner function of a given system represents its most complete characterization and enables
predicting the result of any possible measurement one may perform on the system \cite{qse}. The results
indicate the existence of strong correlations between the two spatial/temporal modes carrying a
coherently delocalized single photon. Whether or not such correlations are nonlocal cannot be asserted;
for instance, the Bell's inequality test performed in Ref.\cite{lvovsky_prl04_2} is based upon
dichotomization of quadrature data and heavily relies upon data-rejection. Of particular interest, in
this respect, is the experiment by Hessmo et al. \cite{hessmo_prl04}, who have implemented the
theoretical proposals by Tan et al. \cite{tan_prl91} and by Hardy \cite{hardy_prl94}, demonstrating the
nonclassical character of the correlations characterizing two spatial modes sharing a single photon.

Similar to the two-photon case, also ``entangled" single-photon sources may find immediate application
in quantum information technology: single-photon ebits have indeed been proven to enable linear optics
quantum teleportation \cite{ent_1phot,gisin_tele} and play a central role in linear optics quantum
computation \cite{knill,pittman}. Furthermore, the time-bin entanglement of the kind of Ref.\cite{noi}
(see also \cite{gisin,simon}) has been proven suitable for long distance applications
\cite{gisin_crypto,gisin_tele}, where the insensitivity to both depolarization and polarization
fluctuations becomes a strong
requirement. 

In this paper, we wish to make one step further in the direction of understanding and proving single
photon N-mode entanglement. To this end, we employ the single-photon source presented in Ref.\cite{noi}
(see Sect.\ref{sec:setup}), perform two-mode homodyne tomography (Sect.\ref{sec:exp_res}) and employ the
reconstructed two-mode density matrix and Wigner function to quantify the degree of entanglement (Sect.
\ref{ssec:entangl}) and implement a novel kind of tomographic Bell's inequality test (Sects.
\ref{sec:ineq_th} and \ref{sec:bell}). In particular, we evaluate the degree of entanglement
characterizing our single-photon system by employing Vidal and Werner's reformulation \cite{vidal02} of
Peres separability criterion \cite{peres96}. Then, we employ the experimentally reconstructed two-mode
Wigner function to perform an innovative continuous variable Bell's inequality test, based on the
theoretical proposal by Banaszek and Wodkiewicz \cite{banaszek}, which we briefly introduce in the next
section.

\section{Tomographic test of Bell's inequality: basic idea}\label{sec:ineq_th}

Let us start by briefly reviewing the theoretical proposal of Ref.\cite{banaszek}: the strict connection
between the displaced parity operator (a dichotomic observable) and the Wigner function
\cite{leonhardt}, suggests that the Wigner function of any two-mode/two-particle system (whether
positive or negative) may play the role of a nonlocal correlation function. Bell's inequality can thus
be recast in the form:
\begin{equation}\label{ineq}
|\mathcal{B}|=\frac{\pi^2}{4}|W(0,0)+W(\alpha_1,0)+W(0,\alpha_2)-W(\alpha_1,\alpha_2)| < 2,
\end{equation}
where $W(\alpha_1,\alpha_2)$ is the value of the two-mode Wigner function at the phase-space point
($\alpha_1=x_1+iy_1,\alpha_2=x_2+iy_2$) defined by the quadratures $x_i$ and $y_i$ of the $i$-th
mode/particle ($i={1,2}$). This inequality, which we shall name Banaszek-Bell's inequality, applies both
to two-particle and to two-mode entangled systems, including spatially/temporally delocalized single
photons. Interestingly, compared to all Bell's inequalities theoretically proposed for continuous
variables and based upon a-posteriori dichotomization of the measured field quadratures
\cite{bell_dichot}, Banaszek-Bell's inequality may achieve higher levels of violation. For instance,
based on the predictions of quantum mechanics, the time-bin encoded single-photon state \cite{noi}:
\begin{equation}\label{state_eq}
|\Psi\rangle = \frac{1}{\sqrt{2}}(|1\rangle_n |0\rangle_{n+1} + |0\rangle_n |1\rangle_{n+1}),
\end{equation}
where $n$ denotes a well-defined temporal mode (or time bin), is expected to maximally violate the
inequality of Eq.~(\ref{ineq}), giving $|\mathcal{B}|\approx 2.2$ for $\alpha_1=\alpha_2\approx 0.3$.

The experimental test of Banaszek-Bell's inequality, in the form written in Eq.~(\ref{ineq}), is
feasible: our idea is to perform two-mode homodyne detection on the two distinct time bins ($n$ and
$n+1$) carrying the temporally-delocalized single photon of Eq.~(\ref{state_eq}), and to use the
experimental data to tomographically reconstruct the two-mode Wigner function entering the inequality of
Eq.~(\ref{ineq}). Our ultra-fast time-domain homodyne detection scheme has the potential to achieve this
goal \cite{marco_science_04,marco_pra70_04}.

It is certainly true that, different from standard Bell's inequality tests, our tomographic test imposes
an indirect approach to Bell's inequality: rather than directly employing the results of correlation
measurements, our scheme requires the manipulation of the experimental quadrature data in order to
reconstruct the Wigner function entering Eq.~(\ref{ineq}). Note, however, that the reconstruction
procedure is rather transparent, does not imply additional hypotheses on the state under study, and can
only introduce noise, without \textit{hiding} or \textit{enhancing} any information contained in the
data. Then, finding a violation of Eq.~(\ref{ineq}) through a tomographic approach indicates that also
direct measurements would lead to a violation.

\section{Experimental setup}\label{sec:setup}

\begin{figure}[ht]
\includegraphics [width=8.5cm]{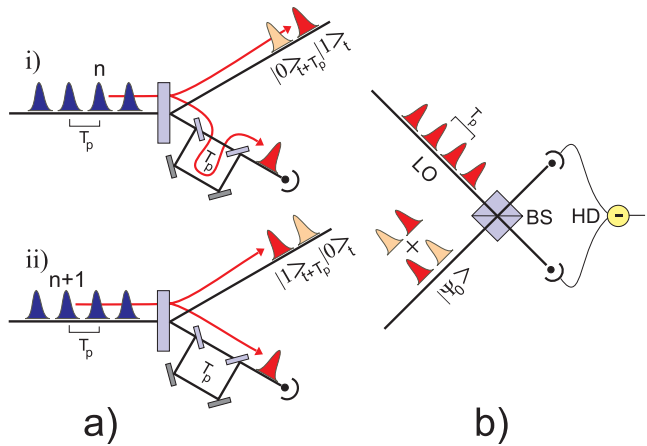}
\caption[] {\label{fig_scheme} (color online). a) Scheme for the remote preparation of a temporally
delocalized single photon; i) and ii) indicate the two indistinguishable alternatives leading to the
coherent superposition of Eq.~(\ref{state_eq}) in the signal arm. b) Scheme for the two-mode,
time-domain, homodyne tomography of such a time-delocalized single-photon state.}
\end{figure}

Let us now consider the experimental setup: a mode-locked Ti:sapphire laser emitting $1.5$ ps pulses at
$786$ nm at a repetition rate of $82$ MHz, is frequency doubled in a LBO crystal; the resulting train of
phase-locked pulses impinges on a nonlinear BBO crystal, cut for degenerate non-collinear type-I SPDC
\cite{spdc_pulse}. Signal-idler photon pairs centered around $786$ nm are generated in symmetric
directions. Before entering an unbalanced fiber-based interferometer and being detected by a
single-photon counter, the idler (trigger) photons undergo a narrow spectral and spatial selection
aiming to the conditional generation of a pure single-photon state in the signal mode (see, e.g.,
\cite{lvovsky_1phot,marco_pra70_04}). Our setup is characterized by a single-photon preparation
efficiency $\eta_{p}= .85$, which depends on both the dark counts of the trigger detector and on the
purity of the prepared single photon \cite{note_purity}. The conditionally-prepared signal beam impinges
on a 50-50 beam splitter (BS) where it is mixed with a strong local oscillator (LO) made of an
attenuated version of the train of laser pulses. Two-mode, high-frequency, time-domain, balanced
homodyne detection (HD) is then performed on two consecutive signal pulses, as depicted in
Fig.~\ref{fig_scheme}b). The detection efficiency ($\eta_{d}= 0.74$) depends both on the efficiency of
the two photodiodes ($\eta_{PD}=.88$) and optical losses, and on the mode-matching with the local
oscillator ($\eta_{MM}= .86$). The overall experimental efficiency of our setup is thus expected to be
of the order of: $\eta =\eta_p \eta_d=.63$.

From the preparation viewpoint, the key part of our setup is the interferometer inserted in the idler
channel: by setting the interferometer delay $T$ equal to the fixed inter-pulse separation $T_p$
characterizing the train of phase-locked pump pulses, a click in the idler channel does not distinguish
the idler photon generated by the $n^{th}$ pump pulse which travelled the long arm, from the idler
photon generated by the $(n+1)^{th}$ pump pulse which travelled the short arm (see
Fig.~\ref{fig_scheme}a)). For an infinite train of mode-locked pump pulses and in the case of equal
losses in the two arms of the interferometer, a click in the trigger detector projects the signal single
photon onto the coherent superposition of Eq.~(\ref{state_eq}).

\section{Experimental results}\label{sec:exp_res}

In order to check the degree of entanglement and to perform the tomographic test of Bell's inequality on
such conditionally and remotely prepared temporally-delocalized single photons, we have performed
two-mode homodyne detection on the pair of consecutive signal time bins sharing the single photon; this
is pictorially drawn in Fig.~\ref{fig_scheme}b). By making use of what we have recently named ``remote
homodyne tomography'' \cite{noi} to remotely vary the relative phase between the two-mode state and the
LO pulses, we have performed about $10^6$ quadrature measurements on the two consecutive signal temporal
modes carrying the delocalized single photon. This has been made possible by the ultrafast operation of
the homodyne detector recently developed in our laboratory~\cite{marco_pra70_04,marco_science_04}.

The measured field quadratures have been employed to reconstruct the density matrix of the measured
system by means of quantum tomography. In particular, we employ the pattern function method (PF)
proposed by D'Ariano, et al.~\cite{dariano_den} to directly retrieve the elements of the two-mode
density matrix in the number-state base from the measured quadrature data: $\rho_{klmn} = \langle
k,l|\hat\rho|m,n\rangle $ is obtained from the statistical average of the corresponding pattern
functions $f_{km}(x,\theta)$ over all homodyne data, which is:
\begin{equation}
\rho_{klmn} = \langle f_{km}(x_1,\theta_1)f_{ln}(x_2,\theta_2)\rangle_{x_1,\theta_1,x_2,\theta_2}
\end{equation}
where $x_{1,2}$ are the quadratures measured at phases $\theta_{1,2}$ on the two consecutive time bins.
The pattern functions have the form $f_{km}(x,\theta) = F_{km}(x)e^{-i(m-k)\theta}$, with
$F_{km}(x)=\partial\left(\psi_k(x)\phi_m(x)\right)/\partial x$ ($m\ge k$), where $\psi_k(x)$ and
$\phi_m(x)$ are, respectively, the regular and irregular wave functions of the harmonic oscillator
~\cite{dariano97,leonhardt}.
In the experiment, $\theta_2-\theta_1$ is remotely varied by means of the interferometer while
$\theta_2+\theta_1$ is left random since our system is independent of the global phase.

\begin{figure}[ht]
\includegraphics [width=8cm]{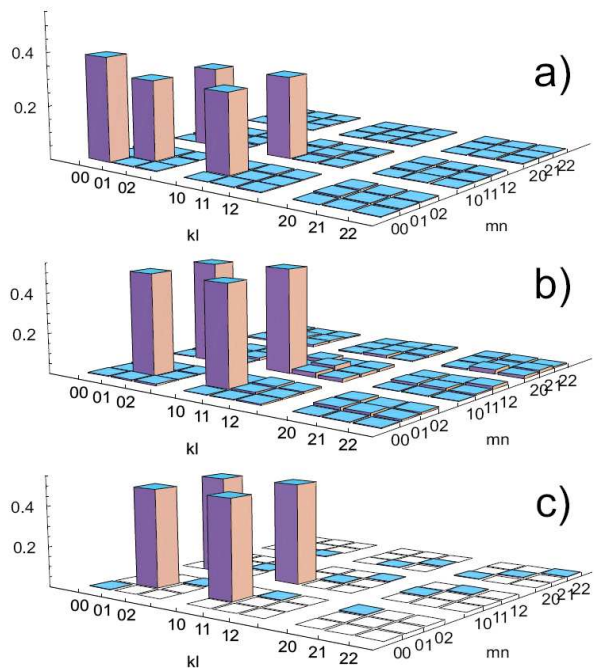}
\caption[] {\label{fig_dens} (color online). a) Experimentally reconstructed density matrix elements
(absolute values) relative to the mixed state of Eq.~(\ref{rho_eq}); b) same elements after ``vacuum
cleaning'' by an inverse Bernoulli transformation; c) density matrix elements reconstructed with the
maximum-likelihood method incorporating the experimental inefficiency.}
\end{figure}

When reconstructing the quantum state, we allow each single-mode Fock state to contain from zero to two
photons; each matrix index is thus varied from zero to two and a total of $3^4=81$ density matrix
elements are reconstructed for the two-mode system, as shown in Fig.~\ref{fig_dens}a. No assumptions are
made on the system to be reconstructed; we just impose a truncation to the reconstruction space since no
multi-photon contributions are expected.

The density matrix of the measured system has also been reconstructed by means of the maximum-likelihood
method (ML) \cite{maxlik}: the density matrix $\hat\rho$ that most likely represents the homodyne data
is retrieved by maximizing a functional $\mathcal{L}(\hat \rho)$ involving the POVM associated with
two-mode homodyne measurements. The only constraint we impose is that the density matrix $\hat\rho$ is a
positive-definite hermitian matrix with unitary trace. In order to limit the number of free parameters
in the minimization procedure, we impose $\hat \rho$ to be independent of the global phase
$\theta_2+\theta_1$. In principle, the ML method enables taking into account the experimental
imperfections, thus reconstructing the density matrix of the measured system corrected for the
experimental efficiency $\eta$. At this stage, we do not impose any such correction.

Both the PF and the ML method (without correcting for $\eta$) give very similar results; hence, in
Fig.~\ref{fig_dens}a, we just plot the elements obtained from the PF method. The reconstructed density
matrix contains both the expected state of Eq.~(\ref{state_eq}) and a vacuum component ($\rho_{0000}$)
coming from the non-unitary experimental efficiency. The measured system is thus given by the mixture
\begin{equation}\label{rho_eq}
\hat{\rho}_s= (1-\eta)|0\rangle_{s \: s} \langle0|+\eta|\Psi\rangle_{s \: s} \langle \Psi|,
\end{equation}
with $|\Psi\rangle_s$ as given in Eq.~(\ref{state_eq}); note that almost no multi-photon contribution
exists in Fig.~\ref{fig_dens}a), as expected. From the vacuum component of the reconstructed density
matrix we evaluate the overall efficiency to be $\eta=0.61$, a result in good agreement with our
estimated single-photon preparation and detection efficiencies.

\subsection{Degree of entanglement}\label{ssec:entangl}

The reconstructed density matrix will now be employed to quantify the amount of entanglement
characterizing the measured two-mode system. To this end, we use the logarithmic negativity
$E_\mathcal{N}$ defined in~\cite{vidal02}, on the line of Peres separability criterion~\cite{peres96}.
In general, Peres criterion gives a {\it necessary} condition for separability \cite{peres96}: if a
bipartite system is separable, the partial transpose of its density matrix (i.e., the transpose
$\hat\rho^{T_A}$ of subsystem $A$ alone) remains a physical density matrix with positive eigenvalues.
Then, if we find that the partial transpose of our density matrix is not positive definite, we know our
single-photon system is certainly non separable. In order to quantify the degree of non-separability
(i.e., the degree of entanglement), one can evaluate the logarithmic negativity defined as
\cite{vidal02}: $E_\mathcal{N}(\hat \rho)\equiv \log_2 ||\hat\rho^{T_A}||_1$, where the symbol $||.||_1$
indicates the trace norm: $||\hat{\sigma}||_1 \equiv Tr \sqrt{\hat \sigma^\dagger \hat \sigma}$, which,
for Hermitian matrices, reduces to $Tr|\hat{\sigma}|= 1+ 2 |\mathcal{N}|$, with $\mathcal{N}$ sum of the
negative eigenvalues of $\hat{\sigma}$ \cite{vidal02}. In other words, the trace norm accounts for the
negative eigenvalues of $\hat\rho^{T_A}$, and is thus directly related to the amount by which the
partial transpose fails to be positive definite. In fact, for a separable system,
$||\hat\rho^{T_A}||_1=1$ and the corresponding logarithmic negativity is zero; for a maximally entangled
$2 \times 2$ bipartite system, $||\hat\rho^{T_A}||_1=2$ and the corresponding logarithmic negativity is
one.

In our case, the reconstructed density matrix yields a logarithmic negativity
$E_\mathcal{N}(\hat\rho)=0.404 \pm 0.001$, thus showing the presence of some amount of entanglement even
for non perfect detection efficiency. This result is perfectly in line with the value expected from
Eq.~(\ref{rho_eq}), with $\eta=0.61$. Interestingly, the bipartite system described by
Eq.~(\ref{rho_eq}) is one of those ``striking" mixtures which remain inseparable (i.e.,
$\hat{\rho}^{T_A}$ has a negative eigenvalue) for any value of the efficiency $\eta$ \cite{peres96}. A
coherently delocalized single photon is thus extremely robust against losses: its degree of entanglement
may decrease but never vanishes.

\subsection{Test of Banaszek-Bell's inequality}\label{sec:bell}

\begin{figure}[ht]
\includegraphics[width=8.5cm]{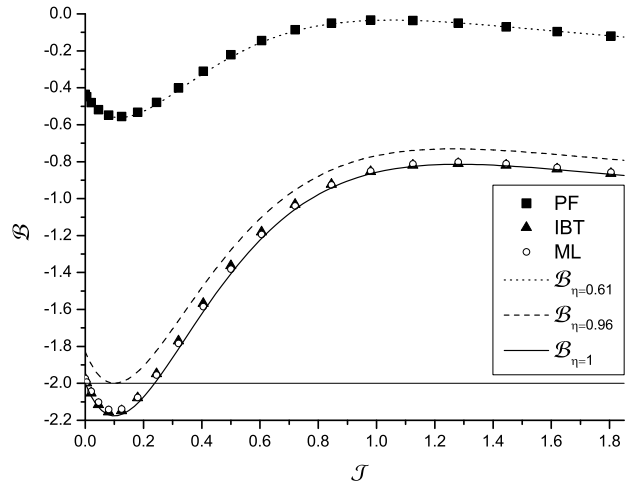}
\caption[] {\label{fig_viola} Plot of the parameter $\mathcal{B}_{\eta}(\mathcal{J})$
(Eq.~(\ref{ineqloss})) for three different values of the overall efficiency $\eta$. Continuous curves
are theoretical predictions, while the points come from the experimental data. The horizontal line
indicates the lower bound imposed by Banaszek-Bell's inequality.}
\end{figure}

Based on the above result, our single-photon two-mode system is characterized by purely quantum
correlations. Our next step is to check whether such quantum correlations are nonlocal. To this end, we
reconstruct the two-mode Wigner function of the measured system \cite{leonhardt}:
\begin{eqnarray}
W(x_1,y_1;x_2,y_2) = \sum_{k,l,m,n} \rho_{klmn} W_{km}(x_1,y_1) W_{ln}(x_2,y_2)
\end{eqnarray}
where  $W_{ij}(x,y)$ is the Wigner function associated with the projector $|i\rangle\langle j|$, and
$\rho_{klmn}$ is the generic element of the experimentally reconstructed density matrix.
The reconstructed Wigner function can now be used to evaluate the combination defined in
Eq.~(\ref{ineq}), thus checking for a violation of Banaszek-Bell's inequality. We find that the
parameter $\mathcal{B}$ falls well within the limits imposed by local hidden-variable theories (see
Fig.~\ref{fig_viola}), a result that may be associated with the low overall experimental efficiency. In
fact, in the case of limited efficiency, the expected Wigner function for the state of
Eq.~(\ref{rho_eq}) takes the form:
\begin{equation}\label{wigner_loss}
W_{\eta}(\alpha_1,\alpha_2) = \frac{4 \left[ 1+2 \eta \left( |\alpha_1 + \alpha_2|^2 -1 \right)
\right]}{\pi^2} \, e^{-2|\alpha_1|^2 -2|\alpha_2|^2};
\end{equation}
hence, taking $\alpha_1=\alpha_2 =\sqrt{\mathcal{J}}$, the combination given in Eq.~(\ref{ineq})
becomes:
\begin{equation}\label{ineqloss}
\mathcal{B}_{\eta}= 1 - 2 \eta + e^{-2 \mathcal{J}}[4 \eta (\mathcal{J} - 1)+ 2]- e^{-4 \mathcal{J}} (8
\mathcal{J} \eta -2 \eta + 1).
\end{equation}
Its behavior is plotted in Fig.~\ref{fig_viola} as a function of the squared amplitude $\mathcal{J}$ for
different values of the global efficiency $\eta$ and it is seen to closely match the experimental
results (squared points) for $\eta=61\%$. It is evident that the current level of experimental
efficiency rules out the possibility of a loophole-free test of Banaszek-Bell's inequality, which would
be attainable only for global experimental efficiencies larger than 96$\%$.

To satisfy such strict requirements we need to introduce some auxiliary assumptions. In this
perspective, it is worth noting that, different from photon counting, homodyne tomography allows one to
explicitly \textit{see} the effect of the experimental inefficiencies on the measured system: the
overall efficiency $\eta$ enters into the reconstructed density matrix (Fig.\ref{fig_dens}a) in the form
of vacuum. In particular, we find that the vacuum component remains exactly the same when no
interferometer is inserted in the idler channel and a single signal photon is perfectly localized within
a given temporal mode. This indicates that the non-unitary efficiency $\eta$ represents our inability of
producing and detecting a pure single-photon state, but, once such imperfect single photon is coherently
delocalized between separate time-bins, the coherence is not further affected by losses. This is
apparent from Fig.~\ref{fig_dens}a), where all the density matrix elements associated with the entangled
state of Eq.~(\ref{state_eq}) have approximately equal weights. In other words, both detection and
preparation inefficiencies give rise to state-independent losses as if a beam splitter with
transmissivity $\eta$ were introduced in the signal path, mixing vacuum with the original state (see
Eq.~(\ref{rho_eq}) and Fig.~\ref{fig_dens} a)). Correcting for our non-unitary efficiencies is thus
similar to making the standard fair-sampling assumption \cite{note_assumpt}.

Based on this reasoning, we have chosen to reconstruct the density matrix of the measured system by
correcting for the non-unitary efficiency; this is expected to remove the vacuum component from the
reconstructed density matrix, which is, to make the $\rho_{0000}$ element almost equal to zero and
rescale the other elements accordingly. We will implement such vacuum removal by employing two different
strategies. Banaszek-Bell's inequality will thus be tested under the fair-sampling assumption or, more
precisely, on the resulting ``vacuum-cleaned'' Wigner function.

In the first case, we account for losses by making an inverse Bernoulli transformation (IBT) of the
reconstructed density matrix; the result is shown in Fig.~\ref{fig_dens}b). The basic idea is to model
the non-perfect experimental efficiency by a beam splitter with transmissivity $\eta$ followed by an
ideal detector: the measured system $\hat\rho_{\rm meas}$ can be seen as the attenuated version of the
system incident on such fictitious beam splitter. Since the
output state of the beam splitter is related to the input state by a Bernoulli transformation, inversion of this transformation gives the state $\hat \rho$ of the ``incident" system, before its mixing with vacuum (i.e., before losses affect it) \cite{kiss}. 

In the second case, we directly reconstruct the loss-free density matrix by adopting the
maximum-likelihood method (ML) while taking into account the experimental efficiency $\eta=0.61$. The
results are shown in Fig.~\ref{fig_dens}c) \cite{note_entanglement}.

As shown in Fig.~\ref{fig_viola}, the Wigner functions reconstructed from both the ``vacuum-cleaned''
density matrices give rise to a good agreement between the experimental data and the theoretical
prediction of Eq.~(\ref{ineqloss}), with unitary efficiency. The temporally-delocalized single photon
thus clearly violates the lower bound of Banaszek-Bell inequality under our version of the fair-sampling
assumption.

Despite the observed violation of Banaszek-Bell's inequality, we cannot claim nonlocality for our
single-photon time-encoded two-mode system. In fact, even in the case of unitary efficiency, our
measuring scheme would still not satisfy the locality hypothesis: quadrature measurements on the two
modes are actually done in sequence, so that a physical signal can in principle be exchanged between the
two co-propagating temporal modes while homodyne measurements are performed on them. The existence of
this possibility makes our Bell's inequality test subject to the so-called locality loophole, which
automatically rules out the possibility of claiming non-locality \cite{note_aspect}. However, this
loophole simply derives from our choice of a temporal delocalization of the single photon and can in
principle be eliminated in future experiments by introducing a fast switch which converts the two
co-propagating temporal modes into two spatially-separated modes; simultaneous measurements could then
be performed on each of the two time-bins by two distant homodyne detectors.

\section{Discussions and conclusions}\label{sec:concl}

Our results demonstrate the feasibility of the tomographic approach to Bell's inequality and may open
the way toward new experiments for testing the foundations of quantum mechanics. Most important, our
experiment sheds some light on the hotly debated concepts of single-photon two-mode entanglement and
nonlocality
\cite{hessmo_prl04,hardy_prl94,tan_prl91,nonloc_1phot,lvovsky_prl04_2,banaszek,vanEnk,powlowsky}. On the
one hand, single-photon two-mode states of the kind described by Eq.(\ref{state_eq}) are known to give
rise to interference upon recombination of the two spatiotemporal modes through an interferometric
scheme; in this perspective, the TWC entanglement between two spatial modes sharing a single photon can
be visualized in terms of wave-particle duality: the interference upon recombination supports the wave
picture of light, while the lack of coincidence counts at the two exits of the entrance beam-splitter
brings to evidence its particle nature. On the other hand, we have shown that our single-photon two-mode
system manifests entanglement, both in terms of Peres-like separability criteria and, under additional
hypotheses, in the more restrictive terms of Banaszek's version of Bell's inequality. The description of
single-photon coherence effects in terms of single-photon N-mode entanglement is thus certainly possible
and meaningful, and could contribute to make the counterintuitive wave-particle duality more intuitively
accessible.

In summary, we have evaluated the degree of entanglement of a single photon coherently delocalized
between two distinct temporal modes and have implemented an innovative test of Bell's inequality based
on quantum homodyne tomography. We have shown that, within some auxiliary assumptions, the correlations
between two well-separated temporal modes sharing a single photon violate Banaszek-Bell's inequality. In
this respect, it is worth emphasizing that none of our additional hypotheses is, in principle,
indispensable. In fact, different from Bell's tests relying on post-selection, an improvement in the
experimental efficiencies would allow us to avoid the fair sampling assumption; indeed, a setup with an
almost perfect single-photon preparation efficiency is currently possible. Furthermore a slight
modification of the experimental setup would easily eliminate the locality loophole. Thanks to the high
efficiencies achievable by homodyne detectors, out tomographic approach may represent the first step in
the direction of a loophole-free test of Bell's inequality. In this perspective, also the debated issue
concerning the possibility of extending both concepts of entanglement and non-locality to a coherently
delocalized single photon, finds here an interesting preliminary answer.

This work has been performed in the frame of the ``Spettroscopia laser e ottica quantistica'' project of
the Physics Department of the University of Florence and partially supported by Ente Cassa di Risparmio
di Firenze and MIUR, under the PRIN initiative and FIRB contract RBNE01KZ94.


\begin{thebibliography}{}

\bibitem{schro} E.~Schr{\"{o}}dinger, Naturwissenschaften, {\bf 23}:807-812;823-828;844-849, 1935.

\bibitem{epr} A. Einstein, B. Podolsky, and N. Rosen, Phys. Rev. {\bf 47}, 777 (1935)

\bibitem{bell} J.S. Bell, Physics \textbf{1}, 195 (1964)

\bibitem{bohm} D.~Bohm, {\em Quantum Theory}, Prentice-Hall, Englewood Cliffs, NJ, 1951

\bibitem{note_assumpt}
The most common auxiliary assumptions involved in the experimental tests of Bell's inequality are: 1)
the fair-sampling assumption, namely, the hypothesis that undetected events coming from low detection
efficiencies would follow the same statistics of the measured data set, 2) post-selection, which
consists in disregarding part of the state of a physical system by not measuring it (i.e., imposing a
selective, or state-dependent, loss mechanism) 3) the locality-loophole, which arises whenever a
physical signal could be exchanged between the two systems under investigation during the measurement
process.

\bibitem{qu_infor} C.H. Bennett, {\em et al.}, Nature {\bf 404}, 247 (2000)

\bibitem{tan_prl91} S.M. Tan, D.F. Walls, and M.J. Collett, Phys. Rev. Lett. \textbf{66}, 252 (1991)


\bibitem{nonloc_1phot} E. Santos, Phys. Rev. Lett. \textbf{68}, 894 (1992); L. Hardy, \textit{ibid.} \textbf{75}, 2065 (1995) and refs. therein; D.M. Greenberger, {\em et al.} \textit{ibid.} \textbf{75}, 2064 (1995); A. Peres, \textit{ibid.} {\bf 74}, 4571 (1995); D. Horne, {\em et al.}, Phys. Lett. A {\bf 209}, 1 (1995); K. Jacobs, {\em et al.}, Phys. Rev. A \textbf{54}, R3738 (1996)

\bibitem{hardy_prl94} L. Hardy, Phys. Rev. Lett. \textbf{73}, 2279 (1994)

\bibitem{banaszek} K. Banaszek, and K. Wodkiewicz, Phys. Rev. A \textbf{58}, 4345 (1998); Phys. Rev. Lett. \textbf{82}, 2009 (1999)

\bibitem{lvovsky_prl04_2} S.A. Babichev, {\em et al.}, Phys. Rev. Lett. \textbf{92}, 193601 (2004)

\bibitem{hessmo_prl04} B. Hessmo, {\em et. al.} Phys. Rev. Lett. \textbf{92}, 180401 (2004)

\bibitem{vanEnk} S. J. Van Enk, Phys. Rev. A {\bf 72}, 064306 (2005)

\bibitem{noi} A. Zavatta, {\em et al.}, Phys. Rev. Lett. {\bf 96}, 020502 (2006)

\bibitem{powlowsky} M. Pawlowski and M. Czachor, quant-ph/0507151, and Phys. Rev. A \textbf{73}, 042111 (2006)

\bibitem{leonhardt} U. Leonhardt, \textit{Measuring the Quantum State of Light}, Cambridge University Press, Cambridge, 1997

\bibitem{qse} M.G.A. Paris and J.~Rehacek (Eds.), {\em Quantum State Estimation} (Springer, Berlin, 2004).

\bibitem{ent_1phot} S. Giacomini \textit{et al.}, Phys. Rev. A {\bf 66}, 030302 (2002)

\bibitem{gisin_tele} H. de Riedmatten \textit{et al.}, Phys. Rev. Lett. {\bf 92}, 047904 (2004)

\bibitem{knill} E. Knill, {\em et al.}, Nature (London) {\bf 409}, 46 (2001)

\bibitem{pittman} T. B. Pittman, {\em et al.}, Phys. Rev. A {\bf 71}, 032307 (2005)

\bibitem{gisin} J. Brendel, {\em et al.}, Phys. Rev. Lett. {\bf 82}, 2594 (1999); H. de Riedmatten, {\em et al.}, Phys. Rev. A {\bf 69}, 050304 (2004)

\bibitem{simon} C. Simon, {\em et al.}, Phys. Rev. Lett. {\bf 94}, 030502 (2005)

\bibitem{gisin_crypto} 
I. Marcikic, {\em et al.}, Nature (London) 421, 509 (2003)











\bibitem{vidal02} G. Vidal and R. F. Werner, Phys. Rev. A {\bf 65}, 032314 (2002)

\bibitem{peres96} A. Peres, Phys. Rev. Lett. {\bf 77}, 1413 (1996)

\bibitem{bell_dichot} B. Yurke, {\em et al.}, Phys. Rev. Lett. {\bf 79}, 4941 (1997); J. Wenger, {\em et al.}, Phys. Rev. A {\bf 67}, 012105 (2003); S. Daffer, {\em et al.}, \textit{ibid} {\bf 72}, 034101 (2005)


\bibitem{marco_pra70_04} A. Zavatta, {\em et al.}, Phys. Rev. A \textbf{70}, 053821 (2004)

\bibitem{marco_science_04} A. Zavatta, {\em et al.}, Science \textbf{306}, 660 (2004), and Phys. Rev. A \textbf{72}, 023820 (2005)

\bibitem{spdc_pulse} T.E. Keller, and M.H. Rubin, Phys. Rev. A {\bf 56}, 1534 (1997); Y.-H. Kim, \textit{et al.}, Phys. Rev. A {\bf 62}, 043820 (2000)

\bibitem{lvovsky_1phot} A.I. Lvovsky, {\em et al.}, Phys. Rev. Lett. {\bf 87}, 050402 (2001)

\bibitem{note_purity} By defining the purity parameter as $P=Tr(\hat{\rho}_s^2)$, we get, in the spectral and the spatial domains, respectively: $P_{spectral} = 1/\sqrt{1+\Delta \omega_i^2 /\Delta \omega_p^2}=.98$, and $P_{spatial} = 1/(1+\Delta \kappa_i^2 /\Delta \kappa_p^2)=.86$, where Gaussian profiles with spectral and spatial widths $\Delta \omega_j^2$ and $\Delta \kappa_j^2$ are assumed
both for the pump power spectrum ($j=p$) and for the filter transmission function in the idler channel
($j=i$). The contribution of purity to the single-photon preparation efficiency is: $\eta_{purity}=
\sqrt{P_{spectral} P_{spatial}}= .92$.

\bibitem{dariano_den} G.~M.~D'Ariano, {\em et al.}, Phys. Rev. A \textbf{50}, 4298 (1994).

\bibitem{dariano97} G. M. D'Ariano in {\em Quantum optics and Spectroscopy of Solids}, edited by T. Hakioglu and A. Shumovsky (Kluwer Academic, Dordrecht, 1997), pp. 175-202.

\bibitem{maxlik} K. Banaszek, {\em et al.}, Phys. Rev. A \textbf{61}, 010304(R) (1999)

\bibitem{kiss} T. Kiss, {\em et al.}, Phys. Rev. A \textbf{52}, 2433 (1995)

\bibitem{note_entanglement} After the vacuum-cleaning procedure, we find $E_\mathcal{N}(\hat\rho)=0.99 \pm 0.01$, which is very close to the unitary value expected for the
pure state of Eq.~(\ref{state_eq}). The amount of entanglement required for observing a violation of
Banaszek-Bell's inequality is $E_\mathcal{N} > 0.942$; as far as the identification of entanglement is
concerned, the more restrictive bounds imposed by Bell's inequality with respect to Peres criterion
appear here explicitly.

\bibitem{note_aspect} In this respect, it is worth reminding that also the first historical Bell's inequality tests were affected by the locality loophole; the problem was solved for the first time by A. Aspect et al. (PRL \textbf{49}, 1804 (1982)).

\end{thebibliography}
\end{document}